\documentclass[conference]{IEEEtran}
\IEEEoverridecommandlockouts
\usepackage{cite}
\usepackage{amsmath,amssymb,amsfonts}
\usepackage{graphicx}
\usepackage{textcomp}
\usepackage{xcolor}
\usepackage{comment}
\usepackage{subfigure}
\usepackage[ruled,vlined]{algorithm2e}
\usepackage{algpseudocode}
\usepackage[hidelinks]{hyperref}
\renewcommand{\algorithmicforall}{\textbf{for each}}
\def\BibTeX{{\rm B\kern-.05em{\sc i\kern-.025em b}\kern-.08em
    T\kern-.1667em\lower.7ex\hbox{E}\kern-.125emX}}
\begin{document}

\title{Moving Virtual Agents Forward in Space and Time
}


\author{\IEEEauthorblockN{Gabriel F. Silva}
\IEEEauthorblockA{\textit{Virtual Humans Lab} \\
\textit{PUCRS}\\
Porto Alegre, Brazil \\
gabriel.fonseca94@edu.pucrs.br}

\\
\IEEEauthorblockN{Carlos G. Johansson}
\IEEEauthorblockA{\textit{Virtual Humans Lab} \\
\textit{PUCRS}\\
Porto Alegre, Brazil \\
carlos@ufcspa.edu.br}

\and
\IEEEauthorblockN{Paulo Knob}
\IEEEauthorblockA{\textit{Virtual Humans Lab} \\
\textit{PUCRS}\\
Porto Alegre, Brazil \\
paulo.knob@edu.pucrs.br}
\\

\IEEEauthorblockN{Soraia R. Musse}
\IEEEauthorblockA{\textit{Virtual Humans Lab} \\
\textit{PUCRS}\\
Porto Alegre, Brazil \\
soraia.musse@pucrs.br}
\and
\IEEEauthorblockN{Douglas A. Schlatter}
\IEEEauthorblockA{\textit{Virtual Humans Lab} \\
\textit{PUCRS}\\
Porto Alegre, Brazil \\
douglas.schlatter01@edu.pucrs.br}

}

\maketitle

\begin{abstract}
This article proposes an adaptation from the model of Bianco 
for fast-forwarding agents in crowd simulation, which enables us to accurately fast forward agents in time. Besides being able to jump from one position to another, agents are able to stay inside their track, it means, the new position is calculated taking into account the original global path the agent would follow, if not being fast-forwarded. Obstacles and other agents around are also taken into account when calculating the new position. In addition, we included a personality aspect on agents, which affect their behaviors and, also, be taken into account when jumping to a future time and space. We conducted some experiments to validate our model, which shows that it was able to indeed fast forward agents from a position to another, in a coherent time, sticking to a given global path while avoiding collisions. Finally, we present a use case, showing that our method can fit inside a "Fog of War" system.
\end{abstract}


\section{Introduction}
\label{sec:introduction}

\footnote{Draft version made for arXiv: \url{https://arxiv.org/}}Since the pioneer work proposed by Thalmann and Musse~\cite{MusseThalmann1997}, many other methods were proposed for crowd simulation, each one with a significant contribution. There are methods that deal with crowds from a microscopic point of view~\cite{Pelechano:2007,Reynolds:1987}, as well methods that deal with a macroscopic point of view~\cite{hughes2002continuum,Treuille:2006} and, even, methods which combine both microscopic and macroscopic simulation strategies~\cite{antonitsch2019bioclouds}. Others explored high dense crowds~\cite{Pelechano:2007,Narain:2009}, heterogeneous behaviors~\cite{Zheng:2016}, navigation control~\cite{Paris2007} and personality traits for agents~\cite{durupinar2009ocean,knob2018simulating}.

Despite the great number of methods proposed for the most varied range of subjects concerning crowd simulation, only very few of them tackled the problem of fast-forwarding a simulation or, in other words, instantly jumping agents from a position A in time X to another position B in time Y, while maintaining a coherent path and with a minimum error. Such a feature is especially useful if performance is taken into account. For example, if a simulation is fast-forwarded from frame 200 to frame 300, this interval of 100 frames does not need to be simulated, relieving computational resources. In their work, Bianco et al.~\cite{clic2016fastForward} proposed a method to estimate the future position of agents in crowd simulations. A Pedestrian Dead Reckoning (PDR) method is used, evaluating the prior positions, velocities, and goals of each agent. The final positions are then estimated based on a global environment complexity factor, that aims to impact the fast-forwarding process, as well as the interaction among agents. Their model was extended by Bianco et al.\cite{clic2017eventTreatment} to allow the inclusion of events (e.g. adding an obstacle) during the jumping period.

In games, movement estimation of agents in virtual environments is a common topic of research, especially regarding computer-controlled units in Real-Time Strategy (RTS) games. Due to "Fog of War" systems (i.e. vision restricted to only the player units), that estimation must be implemented considering a partially observable environment. Hagelback and Johanso~ \cite{hagelback2008fogOfWar} use potential fields to control a bot with imperfect information (i.e. affected by the Fog of War) to explore and navigate the environment. Cho et al.~\cite{cho2016fog} explore the effect of Fog of War system in predictions made by machine learning algorithms. Another common approach includes giving perfect information about the environment to computer-controlled units, allowing the movement estimation to deal with obstacles and terrain deformation hidden by the Fog of War. This approach can be combined with a fast forward method, reducing the computational cost of units navigating through the fog.

Although interesting, the method of Bianco et al.~\cite{clic2016fastForward} has space to improve. The global environment complexity factor is too general, which means that the more complex the environment, the more chances to have a high error rate and, thus, agents jumping to positions far away from where they should be. Therefore, in this work, we aim to adapt the model of Bianco et al.~\cite{clic2016fastForward} to make it more accurate. So, given an initial and final time/frame, each agent is able to teleport from its current position A to a future position B, following their own path and time constraints. Thus, we are not only able to teleport agents, but to avoid collision with obstacles and maintain a coherent path based on the agent current position and its goal. Also, our agents can have personality traits, which impact their behaviors and are, also, taken into account in the fast-forwarding method. Finally, as a use case, we integrate our adapted model with a Fog of War system, placing occluded agents in a "suspended" state until the estimated position is reached or a visible area obstruct its path.

This paper is divided as follows: Section~\ref{sec:related_work} presents the related work regarding position estimation of agents in crowd simulations and real-world pedestrians, along with methods regarding crowd simulation optimizations during movement prediction. Section~\ref{sec:proposed_model} presents the methodology used for integrating a path planning algorithm and personality traits of agents within the fast-forwarding model, as well as how we include personality traits to our agents. Section~\ref{sec:results} presents the results achieved in the integration process and a use case of our proposed method in a fog of war system. Section~\ref{sec:conclusion} presents the final considerations and future work of our method.

\section{Related Work}
\label{sec:related_work}



Due to the increasing demands of simulated environments, such as a higher number of agents and larger scenarios, several methods have been presented to reduce the computational cost of agent movement, collision detection, and collision avoidance on crowd simulations. Pettre et al.~\cite{pettre2006scalableSimulation} presented a method where a level of simulation (LOS), similar to a level of detail (LOD), is given to different sections of the navigation mesh based on the camera viewpoint. Agents closer to the camera are updated with higher frequency. Farther away and occluded agents are given lower priority. Osborne and Dickinson~\cite{osborne2010hierarchicalLOD} presented a similar method for grouping agents using a hierarchical level of detail (LOD). A group can be defined as a set of agents of the same type, goal, and nearby positions. The LOD of a group is defined based on its distance to the camera. 
Guy et al.~\cite{guy2009clearpath} proposed a parallel method for local collision avoidance using the concept of velocity obstacles. Each agent takes into consideration the current velocity of nearby agents to create a set of cones, each containing the possible directions that will cause a collision. After that, the direction of each agent is adjusted seeking a point not included inside the cones, which avoids collision between them.

Different methods and navigation techniques to estimate the position of real-life pedestrians, which were later adapted for virtual agents, are presented in the literature. Beauregard and Haas~\cite{beauregard2006pda} used a pedestrian dead reckoning (PDR) method combined with acceleration sensors to estimate indoor positions. Taia et al.~\cite{taia2017pdaAStar} combines a PDR technique with A* path planning algorithm~\cite{hart1968formal} to provide a more precise approximation of pedestrian routes. 
Yi et al.~\cite{yi2015pedestrianTravelTimeEstimation} proposed a method for estimating the travel time for pedestrians in crowded scenes. Their model defines a group of regions of interest (ROI) for pairings of both sources and destinations, calculating the traffic flow and densities for each region. Environmental elements and stationary persons are taken into account as obstacles. Abnormal behaviors, such as wandering pedestrians, can be identified based on the deviation from their estimated travel time.

Bianco et al.~\cite{clic2016fastForward} presented a method for estimating the future individual position of virtual agents in crowd simulations. Their work uses a PDR method to define the prior positions of each agent, taking into consideration their goal and speed at the frame where the jump occurs. Future positions are adjusted based on a global environment complexity factor and interaction with other agents. 
This model was later extended by Bianco et al.~\cite{clic2017eventTreatment} to allow the inclusion of events during the time jump period. Events are defined as changes in the environment (i.e. adding, moving of removing obstacles and goals). Also, a metric for comparing crowds was presented, taking into consideration the local densities of uniform regions in the environment to define a global error estimation. 


Differing from methods that focus on collision detection optimization or levels of detail, 
we propose a method that integrates the Pedestrian Dead Reckoning (PDR) position estimation with a global path planning algorithm, allowing us to simulate virtual agents that are aware of obstacles in the environment, including during the fast-forwarding. We also included personality traits in agents that affect their behavior during the simulation, aiming to evaluate their impact on the path planning algorithm and the fast forward process.

\section{Proposed Model}
\label{sec:proposed_model}

\begin{figure}[!ht]
  \centering
  \includegraphics[width=0.5\textwidth]{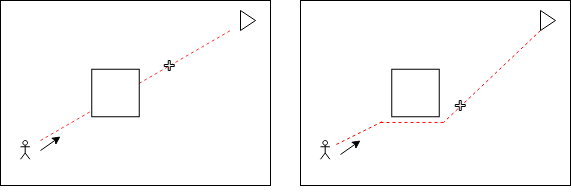}
    \caption{Comparison between the model presented by Bianco et al~\cite{cho2016fog} (left) and our proposed method (right). An agent is placed on the bottom-left corner of the environment, aiming to reach its goal (triangle). The square represents an obstacle obstructing the agent's path. The position of the agent after the fast forward is represented by a plus sign. Our method introduces a global path planning algorithm, that adjusts the agent's path based on obstructing obstacles, in order to fast forward the agent position in time and space. 
    }
    \label{fig:modelOverview}
\end{figure}

The main goal of this work is to adapt the model of Bianco et al.~\cite{clic2016fastForward} to make it more accurate, regarding the impact of obstacles and motion trajectories. To do so, we propose to remove the global environment complexity factor, that was computed taking into account the global free space to move in the environment. Instead, we control the jumping using a navigation method. In other words, such a navigation method would deliver the global path each agent should follow to reach its goal. When the agent needs to jump in time, its new position is calculated taking such a path into account. Also, we include a personality factor in our agents, which should influence their behavior and be taken into account when fast-forwarding the simulation. Section~\ref{sec:proposed_model_time_machine} presents our method to fast forward the simulation, while
Section~\ref{sec:proposed_model_personality} presents our method to include personality on agents and how we deal with it when fast-forwarding a simulation.

\subsection{FF Method Adaptation}
\label{sec:proposed_model_time_machine}

As mentioned before, we chose to adapt the model proposed by Bianco et al.~\cite{clic2016fastForward} to make agents instantly jump from one position to another in time. In short, their method is comprised of four main steps: IP (Interaction with People factor), EC (Environment Complexity factor), PDR (Pedestrian Dead Reckoning), and Repositioning in the space. IP step is responsible for checking how the speeds of each agent should be affected by the presence of nearby agents. EC step is responsible for taking into account the environment complexity (e.g. total area of obstacles) when estimating new positions for the agents. PDR step represents the dead reckoning process on the position estimation, i.e. to where agents go only based on Physics. Finally, the Repositioning step is responsible to reallocate agents in space, following their estimated new positions, but avoiding obstacles and occupied spaces. Their model uses the crowd simulation method known as BioCrowds~\cite{de2012simulating} and, thus, we also do.


We made some adaptations in the method proposed by Bianco et al.~\cite{clic2016fastForward}. First, their model relies on an Environment Complexity factor (EC) to comply with the free region and the presence of obstacles in the way. However, the area of an obstacle is taken into consideration even when not obstructing the path of an agent. We chose to remove such a factor from the model. Instead, we made two alterations: the inclusion of a navigation method, and the position estimation within the defined path. 

Biocrowds~\cite{de2012simulating} is a crowd simulation model that aims to control the local navigation of agents (i.e. define each agent's next position within a pre-defined radius). It can, therefore, have problems to define such next position if an obstacle is in the way. To solve this problem, we chose to use a global navigation method alongside BioCrowds. We implemented the method known as A*, proposed by Hart et al.~\cite{hart1968formal}, which is a method to traverse a graph following the shortest path available from an initial to a destiny position. This method also considers obstacles, allowing agents the circumvent them. So, we use A*, to calculate each agent path, and then 
we compute the total magnitude of the jump in a straight line, as proposed by Bianco et al.~\cite{clic2016fastForward}, to have a value of expected traveled distance based on the speed and position of the agent. Then, we simply project this distance value into the global path of the agent, identifying the point with the same traveled distance. As the A* method takes the obstacle into account, the future position is guaranteed as a point out of any obstacle.

Fig.~\ref{fig:modelOverview} presents a comparison between the method present by Bianco et al.~\cite{clic2016fastForward} (left) and ours (right). The previous method establishes a straight line between an agent's position and its goal (triangle). 
When the fast-forwarding is applied, the agent is repositioned to a target point (plus sign).
If this estimated position is inside the area of an obstacle, Bianco et al.~\cite{clic2016fastForward} identify a nearby available space to place the agent. 
The introduction of a global path planning algorithm allows the identification of obstacles beforehand and removes the possibility of placing an agent inside an obstacle.

Algorithm~\ref{alg:ffAdaptation} presents our Fast Forward Adaptation (FFA). For a given simulation, a stop-frame $t$ and a target frame $t+\Delta t$ are defined, where $t$ represents the frame in which the continuous simulation is interrupted and $t+\Delta t$ represents the frame in which the simulation is resumed after the fast-forwarding. The first step is to estimate the ``future" position of an agent ($pos^{i}_{t+\Delta t}$) using the PDR method, taking into consideration its current position ($pos^{i}_{t}$), movement direction, speed and objective (i.e. final goal). The future position is penalized by an IP (Interaction with People) factor, that considers the presence of other agents. We used the Weibull distribution~\cite{clic2016fastForward}. The magnitude of vector between $pos^{i}_{t+\Delta t}-pos^{i}_{t}$ 
is projected in agent $i$ path, identifying the point in space with the same travel distance (following the path). The agent is repositioned at that point and its path planning is updated, removing sub-goals that where ``skipped" during the FFA and identifying the next immediate step towards the final goal.

\begin{algorithm}[ht]
\label{alg:ffAdaptation}
\SetAlgoLined
\KwData{Stop frame ($t$); Target frame ($t+\Delta t$);}
Continuous Simulation stops at frame $t$;

 \algorithmicforall{ agent $i$ at frame $t$:}{\newline
 $pos^{i}_{t+\Delta t}(x^{i}_{t+\Delta t},y^{i}_{t+\Delta t},z^{i}_{t+\Delta t}) 	\leftarrow PDR(x^{i}_{t},y^{i}_{t},z^{i}_{t}),$\newline
 IP($x^{i}_{t+\Delta t},y^{i}_{t+\Delta t},z^{i}_{t+\Delta t}$)\newline
 Compute($|(pos^{i}_{t+\Delta t} - pos^{i}_{t})|$) \newline
 Repositioning($x^{i}_{t+\Delta t},y^{i}_{t+\Delta t},z^{i}_{t+\Delta t}$); \newline
 UpdatePathPlanning($i$); 
 }
 
Resume Continuous Simulation from frame $t+\Delta t$;
\caption{Fast Forward Adaptation}
\end{algorithm}




\subsection{Personality Traits}
\label{sec:proposed_model_personality}

In order to include personality traits to our agents, we chose the OCEAN (Openness, Conscientiousness, Extraversion, Agreeableness, Neuroticism) psychological traits model, proposed by Goldberg~\cite{goldberg1990alternative}, once it is the most accepted model to define the personality of a person.

It is important to understand each OCEAN factor individually, so we can define its influence on agents. Therefore, we list the definition of each factor, in short: i) Openness (O) = [0;1]: reflects the degree of curiosity, creativity and a preference for novelty and variety; ii) Conscientiousness (C) = [0;1]: reflects the tendency to be organized and dependable; iii) Extraversion (E) = [0;1]: reflects the sociability and talkativeness; iv)Agreeableness (A) = [0;1]: reflects the tendency to be cooperative and compassionate with others; and v) Neuroticism(N) = [0;1]: reflects the degree of emotional stability.

Following the previous definition, we defined the influence that such traits have on the behavior of our agents, during the FFA method. To do so, we follow the method proposed by Knob et al.~\cite{knob2018simulating} to create a relationship between OCEAN traits and behaviors, as following defined: 

\begin{itemize}
    \item Walking Speed [1,2]: defined as a function of Extraversion (E), as follows: $\psi = E + 1$, where $\psi$ is the Walking Speed and $E$ is the Extraversion value;
    \item Leadership [0,1]: defined as a function of Extraversion (E) and Neuroticism (N), as follows: $\omega = (W * E) + ((1 - W) . (1 - N))$, where $\omega$ is the Leadership, $W$ is the weight (defined as 0.5), $N$ is the Neuroticism value and $E$ is the Extraversion value;
    \item Impatience [0,1]: defined as a function of Conscientiousness (C), Agreeableness (A) and Extraversion (E), as follows: $\beta = (W_E * f_E) + (W_{AC} * (1 - A)) + (W_{AC} * (1 - C))$, where $\beta$ is the Impatience, $E$ is the Extraversion value, $C$ is the Conscientiousness value, $A$ is the Agreeableness value, $W_E$ is a weight (defined as 0.1), $W_{AC}$ is another weight (defined as 0.45) and $f_E$ can assume two different values: if $E$ is higher or equal 0.5, it is calculated as follows: $f_E = (2 * E) - 1$. Otherwise, it assumes the value 0.
\end{itemize}

These behaviors (i.e. Walking Speed, Leadership, Impatience) are used to define group features among agents. In other words, such groups would present a default behavior based on the features calculated by the OCEAN trait values. Therefore, we define two group features:

\begin{itemize}
    \item Cohesion ($\zeta_g$) [0,3]: defines how much a group $g$ tends to stay together. The more cohesive the group is, the more agents inside it tend to stay close to each other. At the same time, the less cohesive the group is, the more spread agents of such a group can become. It is calculated in function of the Impatience behavior, as follows: $\zeta_g = (1 - \beta) \times 3$;
    \item Desired Speed ($\Psi_g$) [0,1.2]: defines the desired speed for agents inside the group. It is calculated in function of the Walking Speed behavior, as follows: $\Psi_g = 1.2 \times (\psi - 1)$.
\end{itemize}

Moreover, the Leadership behavior is associated with each agent and defines a given agent that acts as a leader for the other agents of the group. Such a decision is made as follows: the Leadership value of each agent is tested against a threshold (empirically defined as 0.9). If any agent is above such threshold, it is chosen as the leader of the group. If more than one agent surpasses the threshold, one of them is randomly chosen. Finally, if no agent can surpass the threshold, the group has no leader. Agents inside the group follow the behavior of the leader if he is present. Thus, if a strong leader is present, agents inside this group ignore their features and assume the features of the leader. For example, if the leader walks faster, agents of the group will also walk faster. 
In short, we link the OCEAN traits to group features which can be easily taken into account for the FFA method. For example, if an agent has a faster pace (defined by Extraversion OCEAN trait), our fast-forwarding method makes this agent "jump" a higher distance, when compared with agents with a lower pace. It happens so because our method predicts the next position of the agent taking into account, besides other factors, the speed of the agent. Therefore, agents with higher speed are going to jump farther than agents with lower speeds.




\section{Results}
\label{sec:results}

This section presents the results achieved by our method. For all tests, we performed five simulations of each case and calculated the mean values of them. Then, errors were computed as defined in (1), as follows: 

\begin{equation}
\label{eq:errorFormule}
  \sum_{a=1}^{n}\frac{ d(BC^a_{t+\Delta t},FFA^a_{t+\Delta t})}{n},
\end{equation} where $BC^a$ and $FFA^a$ represent the positions of agent $a$, respectively in the simulation, without fast forward method, and in our method, at frame $t$. 
$d$ stands for Euclidean distance and $n$ is the total number of agents in the tested case. The error is the average distance that represents the difference between the continuous simulation and our method, in meters.

Section~\ref{sec:results_time_travel} shows the accuracy of the fast forward method in a obstacles-free environment, while Section~\ref{sec:results_time_travel_obs} shows how our method behave when obstacles are present in the environment. Section~\ref{sec:results_time_travel_cultural} shows how our the personality traits can affect FFA, and finally, Section~\ref{sec:results_fog_of_war} shows a use case for our model.

\subsection{Our method: Fast Forward Adaptation (FFA)}
\label{sec:results_time_travel}

We aim to evaluate if the FFA method is working as intended, it means, agents should be able to jump from its positions in time X to a future position in time Y, in the most accurate way possible. To do so, we modeled two different scenarios. First, a 30x30 (900$m^2$) meters scenario with just one goal. Two simulations were run, varying the number of agents (1 agent and 5 agents). Second, a 30x30 meters scenario with two goals. Again, two simulations were run: first, with two groups of 5 agents each; second, with two groups of 10 agents each. Each group wants to reach a different goal (e.g. group 1 wants to reach goal 1 and group 2 wants to reach goal 2). Table~\ref{tlb:simulationParameters} shows all of these simulations. Also, for all simulations, in both scenarios, we ran the same experiment with and without the FFA method. The idea is to check if our model is working as intended (i.e. calculating valid positions for agents in the simulation, where valid positions can be understood as positions nearby the position the agent would be in the simulation without the fast forward method), independently of the number of agents/groups in the simulation and the number of goals present in the environment (the more goals, the more are the chances of the path of the agents to intersect each other). 
It is worth mentioning that during the execution of our tests, we define that agents should be fast-forwarded from frame 600 to frame 1000. We chose this interval because, in the continuous simulation with two goals, it is the range of time where agents from different groups cross/intersect each other.

Table~\ref{tbl:resultsNoObstacles} presents the results of these simulations. The columns Time, AvgSpeed (average speed), AvgAngVar (average variation of the agent's direction along the path), and AvgDist (average distance between agents) are relative of the Continuous Simulation (i.e. without out FFA method). Time FFA is the total time that the simulations with our FFA method took, while Error represents the relative error between the positioning of the agents in the Continuous Simulation and the FFA simulations.

\begin{table}[!ht]
\centering
\caption{Scenarios of the simulations of the FFA method. We vary the amount of agents, the amount of groups and the amount of goals. To compare the results, we also run the four scenarios without our FFA method (Continuous Simulation).}
\begin{tabular}{|c|c|c|c|}
\hline
Simulation ID & \#Agents & \#Groups & \#Goals \\ \hline
1             & 1        & 1        & 1       \\ \hline
2             & 5        & 1        & 1       \\ \hline
3             & 10       & 2        & 2       \\ \hline
4             & 20       & 2        & 2       \\ \hline
\end{tabular}
\label{tlb:simulationParameters}
\end{table}


\begin{table}[ht]
\scriptsize
\centering
\caption{Results of the FFA method scenarios. The columns Time, AvgSpeed, AvgAngVar, and AvgDist are relative of the Continuous Simulation (i.e. without out FFA method). Time FFA is the total time that the simulations with our FFA method took, while Error represents the relative error between the positioning of the agents in the Continuous Simulation and the FFA simulations.}
\begin{tabular}{c|c|c|c|c|cc}
\cline{2-5}
                          & \multicolumn{4}{c|}{Continuous Simulation}                                                                                                                                                                                      & \multicolumn{2}{c}{}                                                                                                                                    \\ \hline
\multicolumn{1}{|c|}{Sim} & \begin{tabular}[c]{@{}c@{}}Time\\ (s)\end{tabular} & \begin{tabular}[c]{@{}c@{}}AvgSpeed\\ (m/s)\end{tabular} & \begin{tabular}[c]{@{}c@{}}AvgAngVar\\ (º)\end{tabular} & \begin{tabular}[c]{@{}c@{}}AvgDist\\ (m)\end{tabular} & \multicolumn{1}{c|}{\begin{tabular}[c]{@{}c@{}}Time FFA \\ (s)\end{tabular}} & \multicolumn{1}{c|}{\begin{tabular}[c]{@{}c@{}}Avg Error\\ (m)\end{tabular}} \\ \hline
\multicolumn{1}{|c|}{1}   & 8.26                                               & 1.49                                                     & 3.24                                                    & -                                                     & \multicolumn{1}{c|}{6.77}                                                    & \multicolumn{1}{c|}{1.85}                                                \\ \hline
\multicolumn{1}{|c|}{2}   & 13.73                                              & 1.37                                                     & 19.75                                                   & 0.41                                                  & \multicolumn{1}{c|}{11.32}                                                   & \multicolumn{1}{c|}{1.96}                                                \\ \hline
\multicolumn{1}{|c|}{3}   & 15.53                                              & 1.13                                                     & 2.64                                                    & 0.54                                                  & \multicolumn{1}{c|}{12.03}                                                   & \multicolumn{1}{c|}{1.17}                                                \\ \hline
\multicolumn{1}{|c|}{4}   & 24.84                                              & 1.26                                                     & 16.98                                                   & 0.44                                                  & \multicolumn{1}{c|}{19.56}                                                   & \multicolumn{1}{c|}{1.20}                                                \\ \hline
\end{tabular}
\label{tbl:resultsNoObstacles}
\end{table}

As presented in Table~\ref{tbl:resultsNoObstacles}, it can be observed a difference between the times of the simulations with and without the FFA method. Continuous simulations presented higher values of Time than simulations with the FFA method, it means, when using the FFA method, simulations were able to finish faster. It leads to the belief that the FFA method can, indeed, help simulations to run faster by avoiding to simulate a defined interval of frames/time. 
Besides that, the average error between the positioning of agents in the Continuous Simulation and the FFA method is lower than 2 meters for all cases. Such error is calculated as an average displacement in the positioning of all agents present in the simulation, when compared with a continuous simulation. It means that in our 900$m^2$ environment, agents presented, when fast forwarded, a displacement in their positioning below 2 meters. This demonstrates that our method is capable to reduce simulation time without needing to compensate with a big loss in precision.


\subsection{Fast Forward with Obstacles}
\label{sec:results_time_travel_obs}

\begin{figure}[h]
\centering
\includegraphics[width=0.38\textwidth]{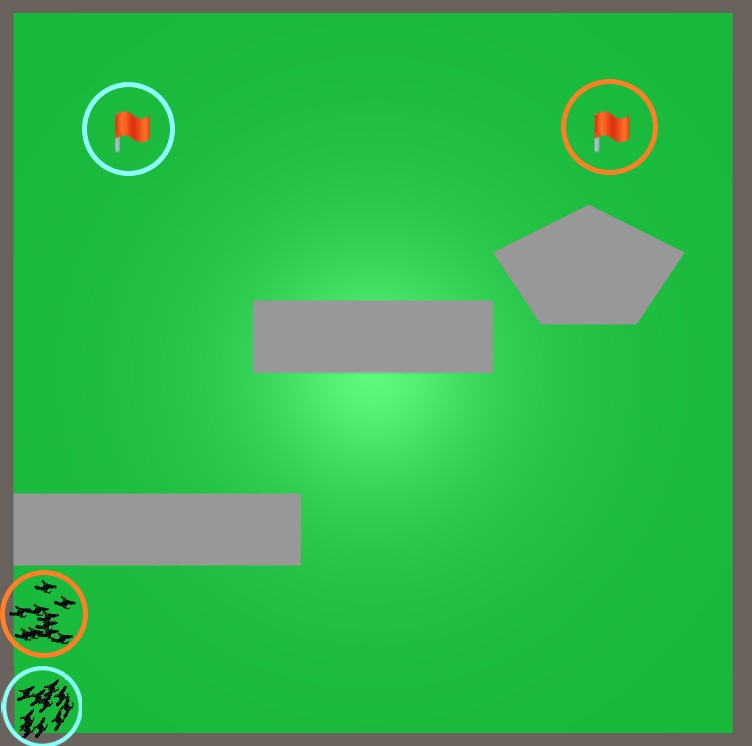}
\caption{Scenario used for the simulations presented in Table~\ref{tlb:simulationParameters}. Obstacles are represented by gray polygon. Goals are represented by flags. Agents are initially place at the bottom-left corner. The agents inside the highlighted circles aims for the goals highlighted in the same color. The goal highlighted in orange was used for the simulations containing only one goal.}
\label{fig:ObsC}
\end{figure}

In this section, we aim to evaluate how our fast forward method behaves with the presence of obstacles in the environment. To do so, we modeled the same two scenarios presented in Table~\ref{tlb:simulationParameters}, but with one difference: we added three obstacles in the environment. The environment setup is shown in Fig.~\ref{fig:ObsC}. We expect that agents are able to jump following their own paths (as defined by the path planning model, explained in Section~\ref{sec:proposed_model_time_machine}) and avoiding collision with obstacles in such path, as explained in Section~\ref{sec:proposed_model_time_machine}.


\begin{table}[ht]
\centering
\caption{Results of simulations with obstacles. Metrics are the same as presented in Table~\ref{tbl:resultsNoObstacles}.}
\begin{tabular}{c|c|c|c|c|cc}
\cline{2-5}
                          & \multicolumn{4}{c|}{Continuous Simulation}                                                                                                                                                                                               & \multicolumn{2}{c}{}                                                                                                                                    \\ \hline
\multicolumn{1}{|c|}{Sim} & \begin{tabular}[c]{@{}c@{}}Time\\ (s)\end{tabular} & \begin{tabular}[c]{@{}c@{}}Avg\\ Speed\\ (m/s)\end{tabular} & \begin{tabular}[c]{@{}c@{}}Avg\\ AngVar\\ (º)\end{tabular} & \begin{tabular}[c]{@{}c@{}}Avg\\ Dist\\ (m)\end{tabular} & \multicolumn{1}{c|}{\begin{tabular}[c]{@{}c@{}}Time FFA \\ (s)\end{tabular}} & \multicolumn{1}{c|}{\begin{tabular}[c]{@{}c@{}}Error\\ (m)\end{tabular}} \\ \hline
\multicolumn{1}{|c|}{1}   & 5.32                                               & 1.31                                                        & 13.81                                                      & -                                                        & \multicolumn{1}{c|}{4.12}                                                    & \multicolumn{1}{c|}{0.25}                                                \\ \hline
\multicolumn{1}{|c|}{2}   & 9.16                                               & 1.49                                                        & 11.29                                                      & 0.36                                                     & \multicolumn{1}{c|}{9.01}                                                    & \multicolumn{1}{c|}{1.18}                                                \\ \hline
\multicolumn{1}{|c|}{3}   & 11.89                                              & 1.37                                                        & 7.92                                                       & 0.45                                                     & \multicolumn{1}{c|}{10.44}                                                   & \multicolumn{1}{c|}{1.23}                                                \\ \hline
\multicolumn{1}{|c|}{4}   & 27.51                                              & 1.36                                                        & 11.22                                                      & 1.36                                                     & \multicolumn{1}{c|}{18.92}                                                   & \multicolumn{1}{c|}{1.87}                                                \\ \hline
\end{tabular}
\label{tbl:resultsObstacles}
\end{table}



Table~\ref{tbl:resultsObstacles} shows the results achieved in the simulations with obstacles. As it was already observed in Table~\ref{tbl:resultsNoObstacles}, continuous simulations presented higher values of Time than simulations with the FFA method. It shows that obstacles are having little or no impact on the behavior of the FFA method. Moreover, Error values were kept below 2 meters, as it was already observed in Table~\ref{tbl:resultsNoObstacles}. It is another evidence that the FFA method can deal with obstacles with little or no impact on the path of the agents.


\subsection{Fast Forward with Personality Traits}
\label{sec:results_time_travel_cultural}

In this section, we aim to evaluate if the fast forward method is being influenced by the personality of the agents. To do so, we modeled the scenario of Simulation 3 presented in Table~\ref{tlb:simulationParameters}, with the three obstacles used in Section~\ref{sec:results_time_travel_obs}. We ran four simulations with fast forward to check for personality variations in groups and we add to our tests the difference data (Diff) in relation to the FFA method with obstacles on the same scenario of simulation, so that we can see the differences in the positions of the agents in the simulations and demonstrate the effect of personalities within the proposed model, as presented in Table~\ref{tab:simulation-scenarios_personality}. In simulations 1 and 2, we tested the Leadership behavior of the group, while in simulations 3 and 4 we tested the Impatience behavior. OCEAN traits values were defined following the behavior definitions explained in Section~\ref{sec:proposed_model_personality}.
For simulations 1 and 2, we expect that groups with a strong leader (Simulation 1) are able to reach their goals faster when compared with groups without a strong leader (Simulation 2), because all agents of the group should follow the leader. Also, since Extraversion OCEAN trait is important to define the mean speed of the agents of a group, high values of this trait reflect in higher velocities as seen in \ref{sec:proposed_model_personality}. 
For simulations 3 and 4, we expect that groups with higher levels of Impatience (Simulation 3) present a more disordered behavior when compared with groups with lower levels of Impatience (Simulation 4). Besides that, we expect that groups with high Impatience present slightly higher speeds.

\begin{table}[!ht]
\centering
\caption{Scenarios for the simulations with personality traits. O, C, E, A and N represent the five traits of OCEAN.}
\begin{tabular}{|c||c||c||c||c||c||c|}
\hline
Simulation & O & C & E & A & N & Test\\ \hline
1 & 0.5 & 0.5 & 0.8 & 0.5 & 0.8 & Leadership  \\ \hline
2 & 0.5 & 0.5 & 0.2 & 0.5 & 0.8 & W/o Leadership\\ \hline
3 & 0.5 & 0.8 & 0.2 & 0.8 & 0.5 & Impatient\\ \hline
4 & 0.5 & 0.2 & 0.8 & 0.2 & 0.5 & Patient\\ \hline
\end{tabular}
\label{tab:simulation-scenarios_personality}
\end{table}


Table~\ref{tbl:resultsOcean} shows the total simulation times for all four simulations. It is possible to notice that the total time of the simulation with a high value of Leadership (Sim 1) is slighter lower than the total time of the simulation with a low value of Leadership (Sim 2). It seems to confirm what we expected: agents inside a group with a strong leader tend to follow its lead and walk at a quicker pace, arriving at their goals faster. Moreover, it can be seen that the total time of the simulation with a high value of Impatient (Sim 3) is greatly lower than the total time of the simulation with a low value of Impatient (Sim 4). Although we did not expect to find so great of a difference, it also seems to confirm what we expected: impatient agents walked at a quicker pace than patient agents, trying to arrive at their respective goals as soon as possible.

\begin{table}[ht]
\centering
\caption{Results of the simulations with personality traits. Metrics are the same as presented in Table~\ref{tbl:resultsNoObstacles}, except for Diff, which represents the difference in agents' position between the simulations with only the FFA method, and simulation with the FFA method along with personality traits.}
\begin{tabular}{c|c|c|c|c|cc}
\cline{2-5}
                          & \multicolumn{4}{c|}{Simulation with OCEAN}                                                                                                                                                                                              & \multicolumn{2}{c}{}                                                                                                                                    \\ \hline
\multicolumn{1}{|c|}{Sim} & \begin{tabular}[c]{@{}c@{}}Time\\ (s)\end{tabular} & \begin{tabular}[c]{@{}c@{}}Avg\\ Speed\\ (m/s)\end{tabular} & \begin{tabular}[c]{@{}c@{}}Avg\\ AngVar\\ (º)\end{tabular} & \begin{tabular}[c]{@{}c@{}}Avg\\ Dist\\ (m)\end{tabular} & \multicolumn{1}{c|}{\begin{tabular}[c]{@{}c@{}}Time FFA \\ (s)\end{tabular}} & \multicolumn{1}{c|}{\begin{tabular}[c]{@{}c@{}}Diff\\ (m)\end{tabular}} \\ \hline
 \multicolumn{1}{|c|}{1}   & 42.56                                               & 0.44                                                        & 10.59                                                      & -                                                        & \multicolumn{1}{c|}{4.12}                                                    & \multicolumn{1}{c|}{6.00}                                                \\ \hline
\multicolumn{1}{|c|}{2}   & 151.49                                               & 0.22                                                        & 24.90                                                      & 0.82                                                     & \multicolumn{1}{c|}{9.01}                                                    & \multicolumn{1}{c|}{10.62}                                                \\ \hline
\multicolumn{1}{|c|}{3}   & 21.76                                              & 1.03                                                        & 30.05                                                       & 1.24                                                     & \multicolumn{1}{c|}{10.44}                                                   & \multicolumn{1}{c|}{2.24}                                                \\ \hline
\multicolumn{1}{|c|}{4}   & 107.92                                              & 0.25                                                       & 27.58                                                      & 0.71                                                     & \multicolumn{1}{c|}{18.92}                                                   & \multicolumn{1}{c|}{11.07}                                                \\ \hline
\end{tabular}
\label{tbl:resultsOcean}
\end{table}


It is important to highlight that simulations 2 and 4 of Table~\ref{tbl:resultsOcean} takes more than double the execution time (i.e. number of frames required for every agent achieve their goal) of simulations 1 and 3 of the same table. In the case of simulation 2, with the aim of the groups to not have a strong leader it is necessary to lower the expressiveness of the OCEAN status, in this case, 0.2 out of 1.0, 
simultaneously this affects the walking speed characteristic of the agents which influence their mean speed. This can be seen when we compare simulation 1 with simulation 2, in the first agents follow the velocity of it strong leader, this makes the agents at this simulation twice as fast second one, as can been seen in table~\ref{tbl:resultsOcean}, therefore taking more time to reach each group destination goal and increasing simulation time.

In the case of simulation 3 and 4, the degree of impatience of a group determines its cohesion as a group. A high impatience degree culminates in a low cohesion value (i. e a high average distance between agents as seen in simulation 3). The impatience of the agents makes them give more importance for positions near its goal then close from its original group. Furthermore, in order to simulate an impatient behavior, we have to increase the expressiveness of the agents. Mutually, this increases it’s walking speed, and hence increasing agents mean velocity. Simulation 4 has a patience behavior, so with more cohesion, which is visible because has the lowest average distance between all tests, and less walking speed, also detectable with a low average speed, this results in a more simulation time for the same reasons as simulation 2.
The gaps in time between tests also present in the fast-forwarding method, due to it be coherent with the influences of personality behaviors. An example is that agent 0 in the simulation 1, with a strong leader, in the moment of the fast forward is in the mean position (6.62,4.32), while in simulation 2, without a strong leader, the same agent, in the same frame, does the fast forward and get the position (3.05,3.05). This coherence demonstrates the efficiency of the fast forward method coexists with the effects of personality behavior, through adapting the fast forward for each personality.


\subsection{Comparison with the state-of-the-art}
\label{sec:results_comparison_cliceres}

In order to validate the accuracy of our model, we performed a comparison with the model we based our work, which means, the model proposed by Bianco et al.~\cite{clic2016fastForward}. To do so, we ran the experiments presented in ~\cite{clic2016fastForward} with a similar setup, using a single environment of 40x23 meters (920$m^2$) containing the following obstacle configurations: no obstacles; 2 obstacles with 109.71$m^2$ total area; 7 obstacles with 128.68$m^2$ total area; and 4 obstacles with 582.22$m^2$ total area. Also, simulations were run with 8, 80, and 160 agents in total.

\begin{figure}[!ht]
  \centering
  \subfigure[fig:clicEnvironmentA][Scenario containing 7 obstacles.]{\includegraphics[width=0.42\textwidth]{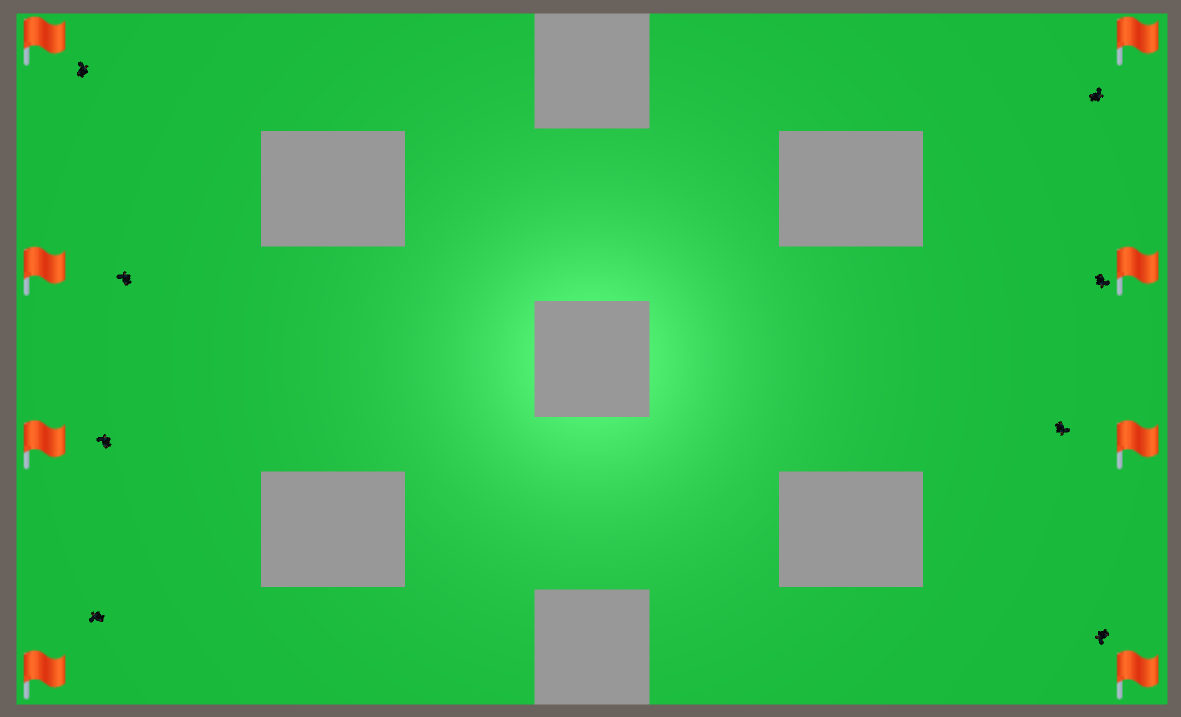}}
  \subfigure[fig:clicEnvironmentB][Scenario containing 4 large obstacles.]{\includegraphics[width=0.42\textwidth]{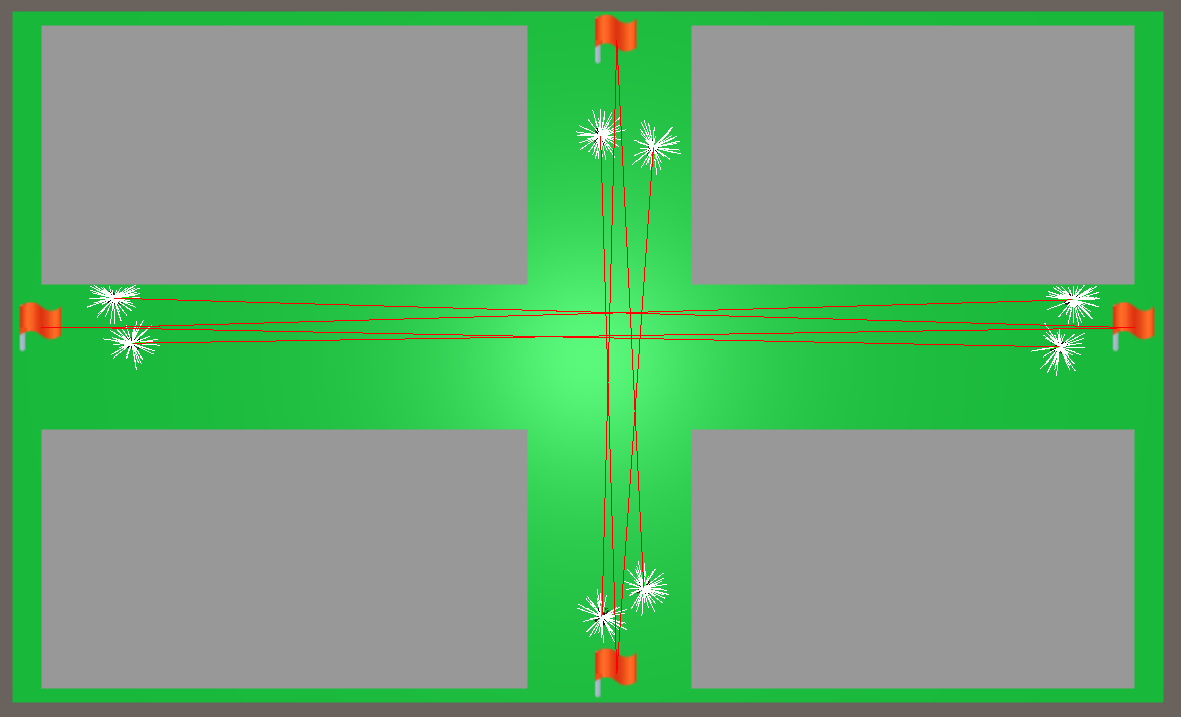}}
    \caption{Scenarios used for comparison with the method presented by Bianco et al.~\cite{clic2016fastForward}. Obstacles are represented by gray rectangles. Goals are represented by flags. In (b), the paths defined for each agent in highlighted in red lines. White lines represent a connection between the center of an agent to their auxins (as presented in the BioCrowds~\cite{de2012simulating} model).}
    \label{fig:clicEnvironmnets}
\end{figure}

Fig.~\ref{fig:clicEnvironmnets}(a) presents the environment used in the 7-obstacle configuration. Agents are initially placed in the horizontal extremities, needing to cross the environment horizontally to achieve their goals (flags). The same setup for agents and goals were used in the simulations containing 2 and no obstacles. Obstacles were intentionally placed in positions where they would obstruct the direct path (i.e. straight line towards the goal) of agents, converging different groups toward similar areas. This setup creates a higher density of agents, especially during moments where opposing groups intersect each other's paths. Fig.~\ref{fig:clicEnvironmnets}(b) presents the environment for the simulation containing 4 obstacles, where the free space is highly reduced. Also, agents are placed in four cardinal extremities of the environment, aiming to reach the opposite extremity. 

\begin{figure*}[!ht]
  \centering
  \includegraphics[width=0.98\textwidth]{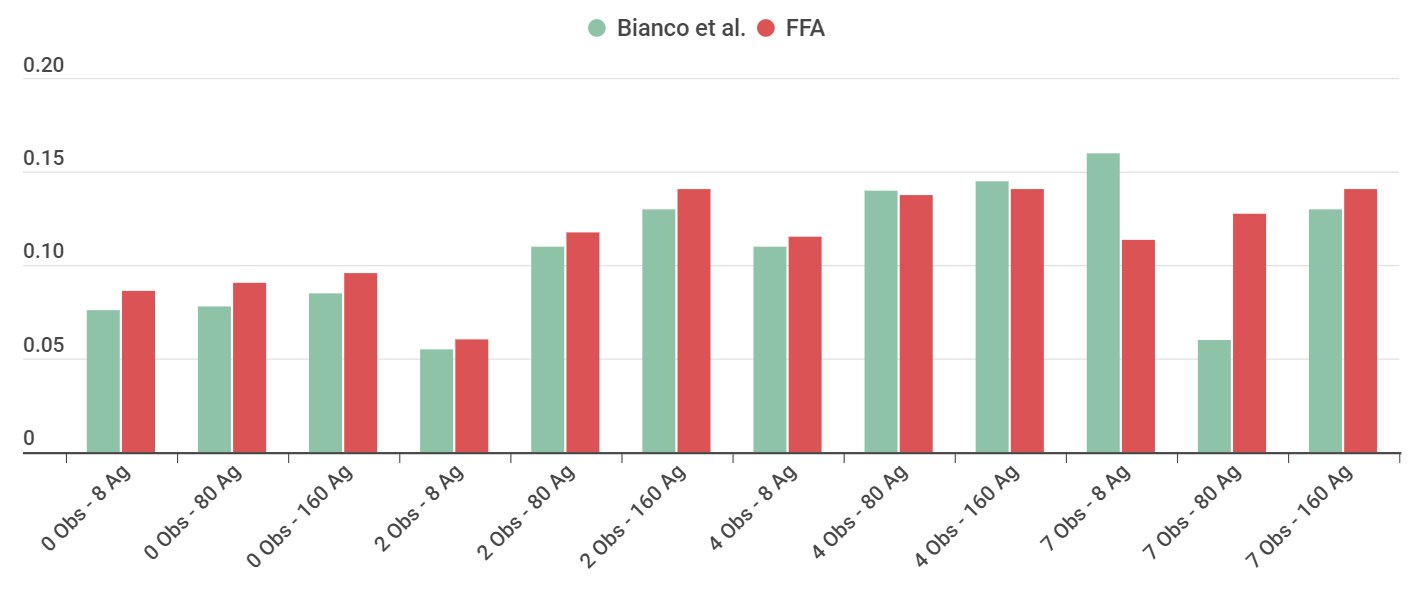}
    \caption{Comparison of relative error in position ($m$) between the model presented by Bianco et al.~\cite{clic2016fastForward} and our proposed FFA method. Values presented are an average between five executions. The value of a single execution is defined by the average $Dif$ of all agents in the simulation, as defined by (2).
    The number of frames affected by the FFA for each scenario is described in Section~\ref{sec:results_comparison_cliceres}.}
    \label{fig:clicComparison}
\end{figure*}

Similar to previous simulations, each scenario was executed five times. Then, we calculate the relative error in position (difference) for each agent, in meters, using the same definition as presented by Bianco et al. ~\cite{clic2016fastForward}. The difference formula is defined in (2) as follows:
\begin{equation}
\label{eq:errorClic}
  Dif^{a}_{t \rightarrow t + \Delta t} = \frac{ d(BC^a_{t+\Delta t},FFA^a_{t+\Delta t})}{d(BC^a_{t},BC^a_{t+\Delta t})},
\end{equation} where $BC^a$ is the position of agent $a$ at a given frame of the continuous simulation. $FFA^a_{t+\Delta t}$ represents the estimated position of agent $a$ at the target frame using our proposed method. $d$ stands for Euclidean distance. The main distinction from (1)
is the inclusion of the distance between the positions of an agent $a$ at the stop frame and the target frame of the simulation. 

Following the definitions of Bianco et al.~\cite{clic2016fastForward}, the continuous simulation in all scenarios where stopped at frame 200 ($t = 200$) and resumed at frame 400 ($t + \Delta t = 400$). The 4-obstacle setup is an exception due to its goal definition. In this scenario, the values $t = 200$ and $t + \Delta t = 370$ were used. The relative error in position ($m$) for a single execution is defined by the average $Dif$ of all agents in the simulation. An average between the five executions is presented in Fig. ~\ref{fig:clicComparison} in comparison with the results achieved by Bianco et al.~\cite{clic2016fastForward}. 

It can be observed that our proposed FFA method achieved a similar error in most scenarios, indicating that the precision of position estimation is maintained with the integration of a global path planning algorithm. The error increases according to the number of agents in the simulation due to the local movement of an agent being affected by nearby agents, making it harder to correctly predict the future position in scenarios with large intersecting groups. Our method maintained this increasing error behavior in all scenarios, differing from Bianco et al.~\cite{clic2016fastForward} in the 7-obstacle setup with 8 and 80 agents. We believe that due to the high number of obstacles in this setup, our path planning algorithm was able to better identify immediate obstructions in the agent's routes, allowing a more precise estimation when compared to Bianco et al.~\cite{clic2016fastForward}.

\subsection{Use Case: Fog of War}
\label{sec:results_fog_of_war}

As a use case, we integrated our method with a simple Fog of War system. In this system, a ``CPU Player" controls the movement of a set of enemy units (i.e agents) in the environment. The ``Human Player" (i.e. user) is then presented with a scenario containing a set of obstacles. Also, the environment is covered in fog, hiding enemy units from the user. The user is also presented with a set of ``watch towers" placed in the environment, revealing areas in the fog. We simulated the behavior of ``CPU" controlled units using our proposed method. The A* algorithm~\cite{hart1968formal} was used to calculate the path of a unit towards a selected goal, and the FFA model was used to estimate the unit's positions within the path at a set of future frames. 

The BioCrowds~\cite{de2012simulating} model uses a grid of cells to represent the space subdivision of the environment. Based on these cells, we create a set of ``fog cells" that cover the environment. Each fog cell hides any enemy unit contained inside its area from the user. Also, a subdivision value ($s$) was used to allow a fog of higher density, therefore, each BioCrowd's cell contained a total of $s^2$ fog cells.
A fog cell contains a state that indicates if it is inside the vision range of a ``watch tower" or a dynamic vision area. Watch towers represent stationary elements (e.g. structures and building of the player) that provides vision of a defined area and reveal a set of fog cells. Dynamic vision areas represent any element that provides vision at different moments of the simulation, also being able to move within the environment (e.g player units). CPU units concealed by the fog (i.e. inside a fog cell that is outside vision) are placed in a ``suspension" state, where no additional calculations are required until this agent leaves such state.

Each fog cell belonging in the agent's path receives a callback that contains the estimated frame where the unit will enter and leave the cell, along with the respective positions at these frames. A callback is activated when the fog cell is within the vision range and the current simulation frame is within the estimated range.
Whenever a callback is activated, the agent leaves the suspension state and is placed back in the environment within the limits of the fog cell. That position is defined using a interpolation between the minimum and maximum estimated position for that cell, taking into consideration the frame of the activation. All remaining callbacks for that agent are deactivated at that point. If the agent leaves vision, it is possible to repeat the process, sending it to the suspension state, and reactivating the callbacks. If the target frame ($\Delta t$) of the FFA is reached without a callback activation, the agent is placed at the final estimated position.

\begin{figure}[!ht]
  \centering
  \subfigure[fig:fogA][Defined path before suspension.]{\includegraphics[width=0.28\textwidth]{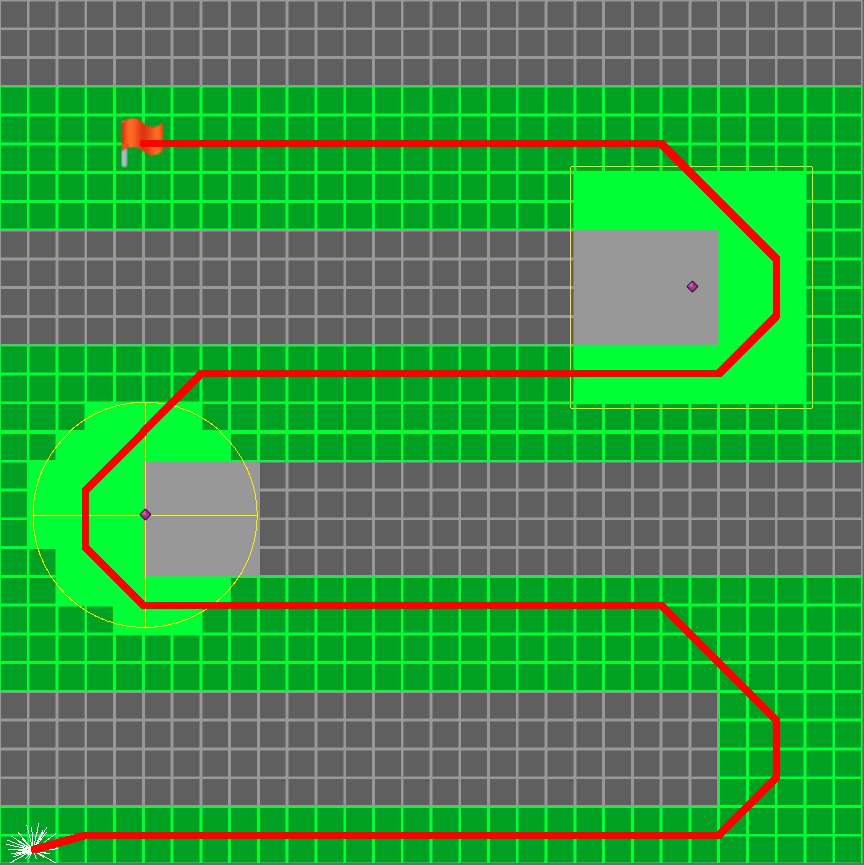}}
  \subfigure[fig:fogB][Suspension interrupted by a revealed area.]{\includegraphics[width=0.28\textwidth]{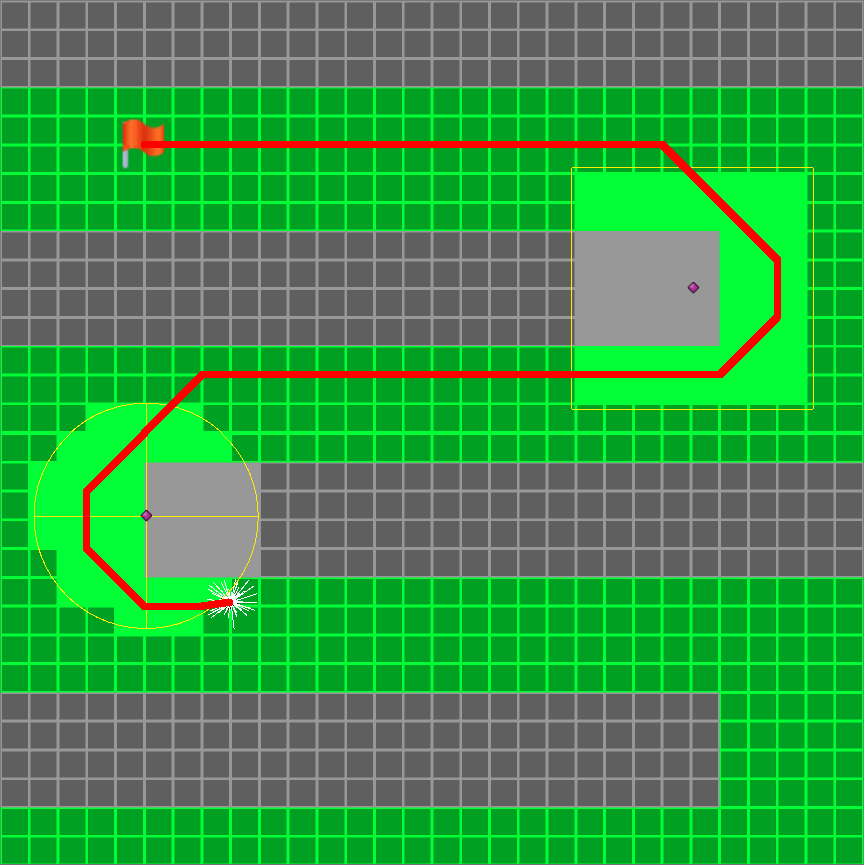}}
  \subfigure[fig:fogC][Suspension not interrupted.]{\includegraphics[width=0.28\textwidth]{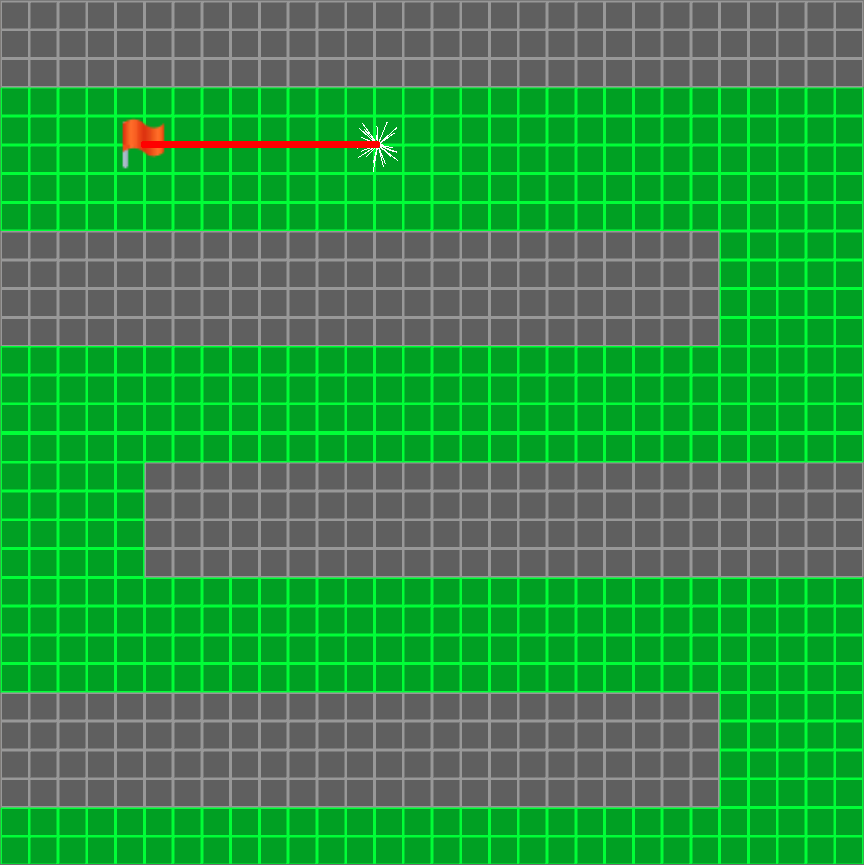}}
    \caption{FFA integration with a Fog of War system. In (a), the defined path of an agent is highlight int red. In (b), the suspension is interrupted due to a ``watch tower" revealing a segment of the path. The agent continues within the path until he leaves vision, entering in another state of suspension. In (c), the set of ``watch towers" is removed, allowing the entire section to be skipped. The agent is then positioned at the predicted point when the target frame is reached.}
    \label{fig:fogofwar}
\end{figure}

Fig.~\ref{fig:fogofwar}(a) presents the setup for the environment, containing 30 x 30 meters (900$m^2$). An agent was placed at the bottom of the scene, aiming to reach a goal (flag) at the top. The path defined by the A* algorithm is highlighted in red, connecting a set of cell centers towards the goal. Two ``watch towers" were placed near the corners of the obstacle: one revealing a circular area; one revealing a square area. The continuous simulation is interrupted at frame 100 ($t = 100$) and the final position is predicted with target frame 3500 ($t + \Delta t = 3500$). Differing from our standard method, the simulation continues from frame 100 onwards. The agent is placed in a suspended state until the frame 3500 is reached or a callback is activated.

In Fig.~\ref{fig:fogofwar}(b), the agent is removed from the suspension state due to a fog cell, that belongs to the path (i.e. contains a callback), being in view-range. The agent is placed in an estimated position for that cell at the current simulation frame, which is around frame 1650 for that environment. At that point, the agent returns to the continuous simulation until it leaves the revealed area, when it enters into suspension state again. In Fig.~\ref{fig:fogofwar}(c), the ``watch towers" are removed and no dynamic vision areas are used. Therefore, all the environment is concealed by the fog of war, and the agent suspension state is not interrupted. In that scenario, the agent enters the suspension state at frame 100 and leaves at frame 3500, close to its goal.

\section{Final Considerations}
\label{sec:conclusion}



This work proposed an adaptation from the model of Bianco for fast-forwarding agents in crowd simulation, which enables us to accurately fast forward agents in time. The future positions of the agents are calculated taking into account the original global path the agent would follow, if not being fast-forwarded. This way, obstacles and other agents around are also taken into account when calculating the new position. Besides that, we included a personality aspect on agents which is taken into account when jumping to a future position.

We ran several tests with our model. The results achieved show that our fast-forwarding method can be a valuable asset for crowd simulation, especially in some given domains. Most notable is the gain in time and resources: since the interval of frames/time where agents are being fast-forwarded does not need to be simulated, the simulation is faster and computational resources can be relieved or spent in other tasks. Also, we integrated our model with a Fog of War system, showing an example of how our model could be allocated in a game.

As for future work, there are several avenues to be followed. We can modify our path planning method (i.e. A*) to consider the number of agents in each cell present in the path of a given agent. It could be used to avoid jams, where a great number of agents become locked by each other. In this case, lanes could be formed by the paths to avoid such jams. Also, in the same way Bianco uses a complexity factor to "punish" the jumping in their method, we could use the complexity of the path to "punish" the jumping in our method. It would help agents to avoid eventual deviation from their original routes (for example, when recalculating their path after the jump). To do both of the cited future works, D* algorithm could be used instead of A*, since it allows dynamic weights for each node of the path. Finally, in our model, we adopted the group behavior from the work of Knob. Although, we did not simulate the fast forward method for groups of agents. Such a feature would allow us to fast forward entire groups of agents (e.g. armies in a game), which would cost less computational resources than to fast forward each agent separately.

\bibliographystyle{IEEEtran}
\bibliography{bib}


\end{document}